%% file: composer.tex
\documentclass[dvipsnames,sigplan,nonacm,10pt]{acmart}
\usepackage{times}
\usepackage{epsfig}
\usepackage{multirow}
\usepackage{float}
\usepackage{tabularx}
\usepackage{enumitem}
\usepackage{hyperref}
\usepackage{xspace}
\usepackage{microtype}
\usepackage{booktabs}
\usepackage[T1]{fontenc}
\usepackage[scaled=0.85]{beramono}
\usepackage{listings}
\usepackage{subcaption}
\usepackage{xcolor}

\newfloat{lstfloat}{htbp}{lop}
\floatname{lstfloat}{Listing}

\newcolumntype{P}[1]{>{\centering\arraybackslash}p{#1}}
\newcolumntype{M}[1]{>{\centering\arraybackslash}m{#1}}

\newcommand{\libsn}{\code{libmozart}\xspace}
\newcommand{\sn}{Mozart\xspace}
\newcommand{\sas}{split annotations\xspace}

\newcommand{\eg}{e.g.,\xspace}

\newcommand{\eat}[1]{}

\newcommand{\one}{{\em (1)}\xspace}
\newcommand{\two}{{\em (2)}\xspace}
\newcommand{\three}{{\em (3)}\xspace}

\newenvironment{denseenum}{
\begin{enumerate}[topsep=1pt, partopsep=0pt, leftmargin=1.5em]
  \setlength{\itemsep}{4pt}
  \setlength{\parskip}{0pt}
  \setlength{\parsep}{0pt}
}{\end{enumerate}}

\renewcommand{\paragraph}[1]{\vspace{1mm}\noindent \textbf{#1}}

\newcommand\code[1]{\lstinline$#1$}

\lstdefinelanguage{AnnotationCDef} {
 keywords={mut, @splittable, unknown, splittype},
 keywordstyle=\color{BrickRed}\bfseries,
 ndkeywords={void, int, long, double, const, MKL_INT, WordSplit, SizeSplit, ReductionSplit, VecSplit, UnknownSplit, MatrixSplit, ReduceSplit, ArraySplit, T1, T2, T3},
 ndkeywordstyle=\color{NavyBlue}\bfseries,
 basicstyle=\small\ttfamily,
 identifierstyle=\color{black},
 sensitive=false,
 comment=[l]{DataFrame},
 morecomment=[s]{/*}{*/},
 morecomment=[l]{\/\/},
 morecomment=[l]{void},
 morecomment=[l]{string},
 morecomment=[l]{vector},
 morecomment=[l]{matrix},
 morecomment=[l]{double},
 commentstyle=\color{darkgray}\ttfamily,
 string=[s]{"}{"},
 showstringspaces=false,
 stringstyle=\color{violet}\ttfamily,
}

\lstdefinelanguage{AnnotationC} {
 keywords={mut, @splittable, unknown, splittype},
 keywordstyle=\color{BrickRed}\bfseries,
 ndkeywords={void, int, long, double, const, MKL_INT, WordSplit, SizeSplit, ReductionSplit, VecSplit, UnknownSplit, MatrixSplit, ReduceSplit, Name, ArraySplit, T1, T2, T3},
 ndkeywordstyle=\color{NavyBlue}\bfseries,
 basicstyle=\small\ttfamily,
 identifierstyle=\color{black},
 sensitive=false,
 comment=[l]{DataFrame},
 morecomment=[s]{/*}{*/},
 morecomment=[l]{\/\/},
 morecomment=[l]{void},
 morecomment=[l]{string},
 morecomment=[l]{vector},
 morecomment=[l]{matrix},
 morecomment=[l]{double},
 commentstyle=\color{darkgray}\ttfamily,
 string=[s]{"}{"},
 showstringspaces=false,
 stringstyle=\color{violet}\ttfamily,
}

\lstdefinelanguage{CustomC}{
 keywords={normalizeMatrixAxis, vdLogp1, vdAdd, vdDiv},
 keywordstyle=\color{BrickRed}\bfseries,
 basicstyle=\small\ttfamily,
 language=C,
 identifierstyle=\color{black},
 sensitive=false,
 comment=[l]{//},
 morecomment=[s]{/*}{*/},
 commentstyle=\color{ForestGreen}\small\ttfamily,
 string=[s]{"}{"},
 showstringspaces=false,
 stringstyle=\color{violet}\ttfamily,
}

\setcopyright{acmlicensed}

\begin{document}

\copyrightyear{2019}
\acmYear{2019}
\acmConference[SOSP '19]{ACM SIGOPS 27th Symposium on Operating Systems Principles}{October 27--30, 2019}{Huntsville, ON, Canada}
\acmBooktitle{ACM SIGOPS 27th Symposium on Operating Systems Principles (SOSP '19), October 27--30, 2019, Huntsville, ON, Canada}
\acmPrice{15.00}
\acmDOI{10.1145/3341301.3359652}
\acmISBN{978-1-4503-6873-5/19/10}

\title{Optimizing Data-Intensive Computations in Existing Libraries with Split Annotations}
\author{Shoumik Palkar and Matei Zaharia}
\affiliation{%
  \institution{Stanford University}
}

\begin{abstract}
\input{abstract}
\end{abstract}

\maketitle

\input{introduction}

\input{overview}
\input{annotations}

\input{client_library}
\input{runtime}
\input{implementation}
\input{integration}
\input{evaluation}

\input{related}
\input{conclusion}

\input{acknowledgements}

\bibliographystyle{ACM-Reference-Format}
\bibliography{composer}

\end{document}

%% file: abstract.tex
Data movement between main memory and the CPU is a major bottleneck in parallel
data-intensive applications. In response, researchers have proposed using
compilers and intermediate representations (IRs) that apply optimizations such
as loop fusion under existing high-level APIs such as NumPy and TensorFlow.
Even though these techniques generally do not require changes to user
applications, they require intrusive changes to the library itself: often,
library developers must rewrite each function using a new IR. In this paper, we
propose a new technique called \emph{split annotations (SAs)} that enables key
data movement optimizations over unmodified library functions. SAs only require
developers to \emph{annotate} functions and implement an API that specifies how
to partition data in the library. The annotation and API describe how to enable
cross-function data pipelining and parallelization, while respecting each
function's correctness constraints. We implement a parallel runtime for SAs in
a system called \sn. We show that \sn can accelerate workloads in libraries
such as Intel MKL and Pandas by up to 15$\times$, with \emph{no} library
modifications. \sn also provides performance gains competitive with solutions
that require rewriting libraries, and can sometimes outperform these systems by
up to 2$\times$ by leveraging existing hand-optimized code.

%% file: introduction.tex
\vspace{-.5em}
\section{Introduction}
\label{sec:introduction}

Developers build software by composing optimized libraries and functions written by other developers. For example, a typical scientific application may compose routines from hand-optimized libraries such as Intel MKL~\cite{mkl}, while machine learning practitioners build workflows using a rich ecosystem of Python libraries such as Pandas~\cite{pandas} and PyTorch~\cite{pytorch}. Unfortunately, on modern hardware, optimizing each library function in isolation is no longer enough to achieve the best performance. Hardware parallelism such as multicore, SIMD, and instruction pipelining has caused computational throughput to outpace memory bandwidth by an order of magnitude over several decades~\cite{stream-benchmark,wulf1995hitting,kagi1996memory}. This gap has made \emph{data movement} between memory and the CPU a fundamental bottleneck in data-intensive applications~\cite{palkar2018evaluating}.

In recognition of this bottleneck, researchers have proposed redesigning software libraries to use optimizing compilers and runtimes~\cite{palkar2017weld,abadi2016tensorflow,kristensen2014bohrium,rocklin2015dask,ragan2013halide,sujeeth2014delite,lee2011implementing,sujeeth2011optiml}. For example, Weld~\cite{palkar2017weld} and XLA~\cite{xla} are two recent compilers that propose rewriting library functions using an intermediate representation (IR) to enable cross-function data movement optimization, parallelization, and JIT-compilation. In both, data movement optimizations such as loop fusion alone have shown improvements of two orders of magnitude in real data analytics pipelines~\cite{palkar2018evaluating,xla-op-fusion}.

Although this compiler-based approach has shown promising results, it
is highly complex to implement. This manifests in two major disadvantages. First,
leveraging these compilers requires highly intrusive changes to
the library itself. Many of these
systems~\cite{palkar2017weld,cyphers2018intel,tvm,xla,kristensen2014bohrium}
require reimplementing each operator in entirety to obtain any benefits. In
addition, the library must often be redesigned to ``thread a compiler'' through
each function by using an API to construct a dataflow graph (e.g., TensorFlow
Ops~\cite{tf-ops} or Weld's Runtime API~\cite{palkar2017weld}). These
restrictions impose a large burden of effort on the library developer and
hinder adoption.  Second, the code generated by compilers might not match
the performance of code written by human experts. For example, even
state-of-the-art compilers tailored for linear
algebra~\cite{ragan2013halide,tvm,xla} generate convolutions and matrix
multiplies that are up to 2$\times$ slower than hand-optimized
implementations~\cite{tvm-mm,halide-mm,tf-slow-ops}. Developers thus face a
tough choice among expanding these complex compilers, dropping their own
optimizations, or forgoing optimization across functions.

In this paper, we propose a new technique called \emph{split annotations}, which provides the data movement and parallelization optimizations of existing compilers and runtimes \emph{without} requiring modifications to existing code. Unlike prior work that requires reimplementing libraries, our technique only requires an \emph{annotator} (e.g., a library developer or third-party programmer) to annotate functions with a split annotation (SA) and to implement a \emph{splitting API} that specifies how to split and merge data types that appear in the library. Together, the SAs and splitting API define how data passed into an unmodified function can be partitioned into cache-sized chunks, pipelined with other functions to reduce data movement, and parallelized automatically by a runtime transparent to the library. We show that SAs yield similar performance gains with up to 17$\times$ less code compared to prior systems that require function rewrites, and can even outperform them by as much as 2$\times$ by leveraging existing hand-optimized functions. 

There are several challenges in enabling data movement optimization and automatic parallelization across black-box library functions. First, a runtime cannot na\"{i}vely pipeline data among all the annotated functions: it must determine whether function calls operating over split data are compatible. For example, a function that operates on rows of pixels can be split and pipelined with other row-based functions, but not with an image processing algorithm that operates over patches (similar to determining valid operator fusion rules in compilers~\cite{xla,palkar2017weld,kristensen2014bohrium,dp-haskell}). This decision depends not on the data itself, but on the shape of the data (e.g., image dimensions) at runtime. To address this challenge, we designed the SAs around a type system with two goals: \one determining how data is split and merged, and \two determining which functions can be scheduled together for pipelining. Annotators use SAs to specify a \emph{split type} for each argument and return value in a function. These types capture properties such as data dimensionality and enable the runtime to reason about function compatibility. For each split type, annotators implement a \emph{splitting API} to define how to split and merge data types.

Second, in order to pipeline data across functions, the runtime requires a lazily evaluated dataflow graph of the annotated functions in an application. While existing systems~\cite{palkar2018evaluating,kristensen2014bohrium,delite,mkl-dnn} require library developers to change their functions to explicitly construct and execute such a graph using an API, our goal is to enable these optimizations without library modifications. This is challenging since most applications will not only make calls to annotated library functions, but also execute arbitrary un-annotated code that cannot be pipelined. To address this challenge, we designed a client library called \libsn that captures a dataflow graph from an existing program at runtime and determines when to execute it with no library modification, and minor to no changes to the user application. We present designs for \libsn for both C++ and Python.
Our C++ design generates transparent wrapper functions to capture function calls lazily, and uses memory protection to determine when to execute lazy values. Our Python design uses function decorators and value substitution to achieve the same result. In both designs, \libsn requires no library modifications.

Finally, once the client library captures a dataflow graph of annotated functions, a runtime must determine how to execute it efficiently. We designed a runtime called \emph{\sn} that uses the split types in the SAs and the dependency information in the dataflow graph to split, pipeline, and parallelize functions while respecting each function's correctness constraints.

We evaluate SAs by integrating them with several data processing libraries: NumPy~\cite{numpy}, Pandas~\cite{pandas}, spaCy~\cite{spacy}, Intel MKL~\cite{mkl}, and ImageMagick~\cite{imagemagick}. Our integrations require up to 17$\times$ less code than an equivalent integration with an optimizing compiler. We evaluate SAs' performance benefits on the data science benchmarks from the Weld evaluation~\cite{palkar2018evaluating}, as well as additional image processing and numerical simulation benchmarks for MKL and ImageMagick. Our benchmarks demonstrate the generality of SAs and include options pricing using vector math, simulating differential equations with matrices and tensors, and aggregating and joining SQL tables using DataFrames. End-to-end, on multiple threads, we show that SAs can accelerate workloads by up to 15$\times$ over single-threaded libraries, up to 5$\times$ compared to already-parallelized libraries, and can \emph{outperform} compiler-based approaches by up to 2$\times$ by leveraging existing hand-tuned code.
Our source code is available at \url{https://www.github.com/weld-project/split-annotations}.

Overall, we make the following contributions:

\begin{denseenum}
\item We introduce split annotations, a new technique for enabling data movement optimization and automatic parallelization with no library modifications.
\item We describe \libsn and \sn, a client library and runtime that use annotations to capture a dataflow graph in Python and C code, and schedule parallel pipelines of black-box functions safely.
\item We integrate SAs into five libraries, and show that they can accelerate applications by up to 15$\times$ over the single-threaded version of the library. We also show that SAs provide performance competitive with existing compilers.
\end{denseenum}

%% file: overview.tex
\section{Motivation and Overview}
\label{sec:overview}

\emph{Split annotations (SAs)} define how to split and pipeline data among
a set of functions to enable data movement optimization and automatic
parallelization. With SAs, an \emph{annotator}---who could be the library
developer, but also a third-party developer---annotates functions and
implements a splitting API that defines how to partition data types in a
library. Unlike prior approaches for enabling these optimizations under
existing APIs, SAs require no modification to the code library developers have
already optimized. SAs also allow developers building new libraries to focus on
their algorithms and domain-specific optimizations rather than on implementing
a compiler for enabling cross-function optimizations. Our annotation-based
approach is inspired by the popularity of systems such as
TypeScript~\cite{typescript}, which have demonstrated that third-party
developers can annotate existing libraries (in TypeScript's case, by adding
type annotations~\cite{typescript-annotations}) and allow other developers to
reap their advantages.

\subsection{Motivating Example: Black Scholes with MKL}

\lstset{language=CustomC}

To demonstrate the impact of SAs, consider the code snippet in
Listing~\ref{listing:bscode}, taken from an Intel MKL implementation of the
Black Scholes options pricing benchmark. This implementation uses MKL's
parallel vector math functions to carry out the main computations. Each
function parallelizes work using a framework such as Intel TBB~\cite{intel-tbb}
and is hand-optimized using SIMD instructions. Unfortunately, when combining
several of these functions on a multicore CPU, the workload is bottlenecked on
\emph{data movement}.

The data movement bottleneck arises in this workload because each function
completes a full scan of every input array, each of which contains \code{len}
elements. For example, the call to \code{vdLog1p}, which computes the
element-wise logarithm of the \code{double} array \code{d1} in-place, will scan
\code{sizeof(double)*len} bytes: this will often be much larger than the CPU
caches. The following call to \code{vdAdd} thus cannot exploit any locality of
data access, and must \emph{reload values for \code{d1} from main memory}.
Since individual MKL functions only accept lengths and pointers as inputs,
their internal implementation has no way to prevent these loads from main
memory \emph{across} different functions. On modern hardware, where computational
throughput has outpaced memory bandwidth between main memory and the CPU by
over an order of
magnitude~\cite{stream-benchmark,wulf1995hitting,kagi1996memory}, this
significantly impacts multicore scalability and performance, even in
applications that already use optimized libraries.

\begin{lstfloat}[t!]
\begin{lstlisting}[language=CustomC]
// inputs are `double` arrays with `len` elements
vdLog1p(len, d1, d1);         // d1 = log(d1)
vdAdd(len, d1, tmp, d1);      // d1 = d1 + tmp
vdDiv(len, d1, vol_sqrt, d1); // d1 = d1 / vol_sqrt
\end{lstlisting}
\captionof{lstlisting}{Snippet from the Black Scholes options pricing benchmark implemented using Intel MKL.}
\label{listing:bscode}
\end{lstfloat}

\begin{lstfloat}[t!]
\begin{lstlisting}[language=AnnotationC]
@splittable(
  size: SizeSplit(size), a: ArraySplit(size),
  mut out: ArraySplit(size))
void vdLog1p(long size, double *a, double *out);

@splittable(
  size: SizeSplit(size), a: ArraySplit(size),
  b: ArraySplit(size), mut out: ArraySplit(size))
void vdAdd(long size, double *a,
           double *b, double *out);
void vdDiv(long size, double *a, 
           double *b, double *out);

\end{lstlisting}
  \captionof{lstlisting}{SAs for three functions in Intel MKL.}
\label{listing:bs-annotations}
\end{lstfloat}

\lstset{language=AnnotationC}

SAs and their underlying runtime, \sn, address this data movement
bottleneck by splitting function arguments into smaller pieces and pipelining
them across function calls. Listing~\ref{listing:bs-annotations} shows the SAs
an annotator could provide for the three functions in
Listing~\ref{listing:bscode}. We describe the SAs fully in
\S\ref{sec:annotations}, but at a high level, annotators use the SA and a new
abstraction called split types to define how to split each function
argument into small pieces. For example, the annotator can define a split type
\code{ArraySplit} to indicate that the array arguments be split into smaller,
regularly-sized arrays. The split type \code{SizeSplit} then indicates that the
size argument be split to represent the lengths of these arrays.  Annotators
then bridge the abstraction of the split type with code that performs the
splitting (e.g., by offsetting into the original array pointer)
by implementing a \emph{splitting API} for each split type.

Given these SAs, \sn executes the code in Listing~\ref{listing:bscode} in a
markedly different way. Instead of calling each MKL function on the full input
arrays at once, \sn splits each array into small, equally-sized chunks that
collectively fit in the CPU cache (e.g., chunks of 4096 elements per array). \sn then
assigns these chunks across threads, and has each thread call all the functions
in the pipeline in sequence (\code{vdLog1p}, \code{vdAdd}, etc.) on one chunk
at a time. Data for each chunk resides in each CPU's local cache across all
functions. Although the total number of loads and stores remains unchanged
(i.e., across all chunks, each function still loads and processes all elements
of \code{d1}, \code{tmp}, and \code{vol_sqrt}), each array element is loaded from main
memory \emph{only once} and served from cache for all subsequent accesses. This
is a stark reduction compared to the original execution, which loads each array
from main memory for each function call.

\begin{figure}[t!]
  \centering
  \includegraphics[width=0.28\textwidth]{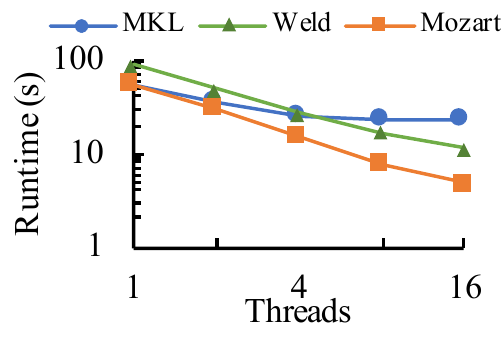}
  \caption{Performance of the Black Scholes benchmark on 1--16 threads with MKL, Weld, and MKL with \sn.}
  \label{fig:blackscholes-motivation}
\end{figure}

Figure~\ref{fig:blackscholes-motivation} shows the impact of SAs on the full
Black Scholes benchmark, which contains 32 vector operations, on a modern Intel
Xeon CPU. The benchmark runs on 11GB of data. While un-annotated MKL bottlenecks
on memory at around four threads, \sn scales to all the cores on the
machine. In this benchmark, \sn also \emph{outperforms} the optimizing Weld
compiler, which applies optimizations such as loop fusion to reduce data
movement by keeping data in CPU registers. We found this was because Weld does
not generate vectorized code for several operators that MKL does vectorize.
Although it is feasible to extend Weld to include SIMD versions of
these operators, this benchmark is one example of the advantages of leveraging
code that developers have already hand-optimized.

\begin{figure}[t!]
  \centering
  \includegraphics[width=0.40\textwidth]{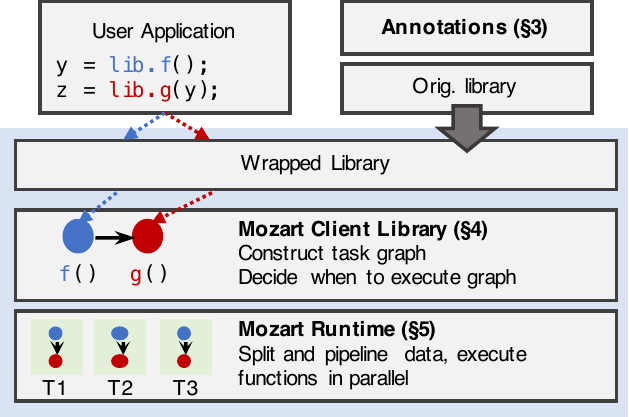}
  \caption{Overview of \sn.}
  \label{fig:overview}
\end{figure}

To enable these performance improvements, \sn first captures a dataflow graph
of annotated functions called in the application (Figure~\ref{fig:overview}).
Since our goal is to leave libraries unmodified, we wish to capture such a
graph without an explicit API that libraries implement.  The \emph{\libsn
client library} (\S\ref{sec:client-library}) transparently handles constructing
such a graph and determining when to execute it, using a combination of
auto-generated wrapper functions and memory protection in C/C++ and function
decorators in Python. \sn then runs a planning step to determine which
functions to pipeline, and then executes tasks using the SAs and
dataflow graph (\S\ref{sec:runtime}).

\subsection{Limitations and Non-Goals}
\label{subsec:limtations}

SAs have a number of restrictions and non-goals. First, since SAs make repeated
calls to black-box functions to achieve pipelining and parallelism, they are
limited to functions that do not cause side effects or hold locks. Second, SAs
do not apply the compute-based optimizations (e.g., common subexpression
elimination or SIMD vectorization) that systems like Weld and TensorFlow XLA
can provide by re-implementing functions in an IR.  Nevertheless, because data
movement has been shown to be one of the major bottlenecks in modern
applications, we show that SAs can provide speedups competitive with these
compilers without rewriting functions.

%% file: annotations.tex
\section{Split Annotation Interface}
\label{sec:annotations}

In this section, we present the split annotation (SA) interface. We first
discuss the challenges in pipelining arbitrary functions to motivate our
design. We then introduce \emph{split types} and \emph{split annotations}, the
core abstractions used to address these challenges. We conclude this section by
describing the \emph{splitting API} that annotators implement to bridge these
abstractions with code to split and merge data.

\subsection{Why Split Types?}
\label{subsec:whysas}

Split types are necessary because some data types in the
user program can be split in multiple ways. As an example, matrix data can
be split into collections of either rows or columns, or can even be partitioned
into blocks. However, some functions require that data be split in a specific
way to ensure correctness. As an example of this, consider the program
below, which uses a function to first normalize the rows of a matrix (indicated
via \code{axis=0}), followed by the columns of a matrix.

\begin{lstfloat}[h]
\begin{lstlisting}[language=CustomC]
normalizeMatrixAxis(matrix, axis=0);
normalizeMatrixAxis(matrix, axis=1);
\end{lstlisting}
\end{lstfloat}

In the first call, the matrix must be split by rows, since the function
requires access to all the data in a row when it is called. Similarly, the
second call requires that the matrix be split into columns. Split types allow
distinguishing these cases, even when they depend on runtime values (e.g.,
\code{axis}). Beyond ensuring correctness, split types also enable pipelining
data split in the same way across functions.

\subsection{Split Types and Split Annotations}
\label{subsec:split-types}
\lstset{language=AnnotationC}

A \emph{split type} is a parameterized type $N\langle V_0 \ldots V_n \rangle$
defined by its name $N$ and a set of parameter values $V_0 \ldots V_n$. A split
type defines how a function argument is split. Two split types are equal if
their names and parameters are equal. If the split types for two arguments are
equal, it means that they are split in the same way, and their corresponding
pieces can be passed into a function together.  \footnote{Since type equality
depends not only on the type name but also on the parameter values, split types
are formally \emph{dependent types}~\cite{dependent-types}.  } Each split type
is associated with a single concrete data type (e.g., an integer, array, etc.).

The library annotator decides what a ``split'' means for the data types in her
library. As an example, an annotator for MKL's vector math library, which
operates over C arrays, can choose to split arrays into multiple
regularly-sized pieces. The annotator can then define a split type
$ArraySplit\langle int, int \rangle$ that uniquely specifies how an array is
split with two integer parameters: the length of the full array and the number
of pieces the array is split into. An array with 10 elements can be split into
two length-5 pieces with a split type $ArraySplit \langle 10, 2 \rangle$, or
into five length-2 pieces with a split type $ArraySplit \langle 10, 5 \rangle$:
these splits have \emph{different} split types since their parameters are
different, even if the underlying data refers to the same array. In the
remainder of this section, we omit the parameter representing the number of
pieces, because \one every split type depends on it, and \two \sn sets it
automatically (\S\ref{sec:runtime}) and guarantees its equality for split types
it compares.

\begin{lstfloat}[t!]
\begin{lstlisting}[language=AnnotationCDef]
 @splittable(
   [mut] <arg1-name>: [<arg1-split-type>|_], ...
 ) [-> <ret-split-type>]
 /* one or more functions */ 
\end{lstlisting}
\captionof{lstlisting}{Full syntax of a split annotation.}
  \vspace{-1em}
\label{listing:sa-syntax}
\end{lstfloat}

A \emph{split annotation (SA)} is then an annotation over a side-effect-free
function that assigns a name and a split type to each of the function's
arguments and its return value. Listing~\ref{listing:sa-syntax} shows the full
syntax of a single SA. For arguments that should not be split (and thus copied
to each pipeline), annotators can give them a ``missing'' split type, denoted
with ``\code{_}'' (e.g.,  ``\code{arg: _}'').  In addition to providing split
types, the SA specifies which of the function arguments are \emph{mutable}
using the \code{mut} tag, which \sn uses to detect data dependencies between
functions when building a dataflow graph (\S\ref{sec:client-library}).
Listing~\ref{listing:bs-annotations} shows SAs for MKL array functions.  Taking
the \code{vdAdd} function as an example, the SA assigns the names \code{size},
\code{a}, \code{b}, and \code{out} to the arguments and assigns each a split
type. The SA marks \code{out} as \code{mut} to indicate that the function
mutates this argument. 

\paragraph{Split Type Constructors.}
One subtlety in writing SAs arises due to the split types' parameters. Even
though a split type's name is known when the annotator writes an SA, its
parameters will generally not be known until \emph{runtime}. For example, in
the $ArraySplit$ split type defined above, the length of an array will not be
known until the program executes. This creates a challenge because \sn
needs to know the full split type including the concrete values of its
parameters.

To address this, a split type can use function arguments to compute its
parameters at runtime.  Specifically, for an SA over a function $F$, each split
type in the SA uses a \emph{constructor} $A_0 \ldots A_n \Rightarrow V_0 \ldots
V_n$ to construct its parameters, where $A_0 \ldots A_n$ are zero or more
arguments of $F$. Within an SA, we use the syntax \code{Name(A0...An)} to refer
to a constructor for a split type with a name \code{Name} that constructs its
parameters with function arguments \code{A0...An}, where \code{A0...An} are
names assigned to arguments in the SA.  Note that the split type constructor is a part of
the \emph{splitting API} that annotators implement for each split type---we
discuss this API further in \S\ref{subsec:splitting-api}.  Unless otherwise
noted, we assume in this paper that split types use the identity function $A_0
\ldots A_n \Rightarrow A_0 \ldots A_n$ as their constructor.

As an example, consider Ex. 1 in Listing~\ref{listing:string-examples}, which
takes a matrix argument and an axis that determines whether the function
operates over rows or columns (similar to the function in
\S\ref{subsec:whysas}). We can represent splitting this matrix by either rows
or columns by using a split type with three integer parameters called
$MatrixSplit \langle int,int,int \rangle$. The parameters represent the matrix
dimensions and the axis to iterate over. Within an SA, an annotator can write
\code{MatrixSplit(m, axis)} to represent this split type:
Listing~\ref{listing:string-examples} shows the constructor definition for this
split type, which maps the matrix and axis into the split type's three
parameters.  The split type for matrices does not depend on the matrix data
itself, since the underlying data does not affect how the matrix is split.  The
SAs in Listing~\ref{listing:bs-annotations} similarly use the \code{size}
argument (but not the array itself) to construct the $ArraySplit$ split type.

With split types, \sn can determine whether two functions can be pipelined
safely. For each annotated function that \sn captures in a dataflow graph, it
initializes the parameters of the split types using the function's arguments.
If all the data passed between two functions have matching corresponding split
types, they can be pipelined. Otherwise, already-split data must be
\emph{merged} and re-split before passing it to the next function to prevent
pipelining errors. Returning to the example from \S\ref{subsec:whysas}, we can
use the split type from Ex. 1 in Listing~\ref{listing:string-examples} to
assign the matrix arguments the split types $MatrixSplit \langle rows,cols,0
\rangle$ and $MatrixSplit \langle rows,cols,1 \rangle$: since these split
types do not match, \sn will not pipeline them.

\begin{lstfloat}[t!]
\begin{lstlisting}[language=AnnotationC]
// Parameters are (rows, cols, axis)
splittype MatrixSplit(int, int, int)
// Constructor for MatrixSplit
MatrixSplit(m, axis) => (m.rows, m.cols, axis)

// Ex. 1: Normalize along an axis in a matrix.
@splittable(mut m: MatrixSplit(m, axis), axis: _)
void normalizeMatrixAxis(matrix m, int axis);

// Ex. 2: Add two matrices element-wise.
@splittable(left: S, right: S) -> S
matrix add(matrix left, matrix right);

// Ex. 3: Scale a matrix element-wise.
@splittable(mut m: S, val: _)
void scaleMatrix(matrix m, double val);

// Ex. 4: Remove zero-valued rows from a matrix.
@splittable(m: S) -> unknown
matrix filterZeroedRows(matrix m);

// Ex. 5: Reduce a matrix to a vector by summing.
@splittable(m: MatrixSplit(m, axis), axis: _) 
  -> ReduceSplit(axis)
vector sumReduceToVector(matrix m, int axis);
\end{lstlisting}
  \captionof{lstlisting}{Examples of SAs over matrices. Ex. 1 shows concrete
  split types, Ex. 2-3 show generics, Ex. 4 shows unknown split types, and
  Ex. 5 shows a reduction function.}
\label{listing:string-examples}
\end{lstfloat}

\paragraph{Generics.}
SAs also support assigning generics to an argument.  Generics in an SA are
similar to generic types in languages such as Java or Rust: if two generics
within an SA have the same name (e.g., \code{S}), the runtime ensures that the
split types they are assigned are equal. Names for generics are local to an SA.
Generics across SAs are propagated via \emph{type inference}
(\S\ref{sec:runtime}), another common feature of existing type
systems~\cite{aiken1993type,pierce2000local}.

Ex. 2 and 3 in Listing~\ref{listing:string-examples} show generics.
Ex. 2 shows a function that adds two matrices element-wise: if \code{left}
and \code{right} are split in the same way (indicated by their matching generic
\code{S}), the function can process them together.

\paragraph{Unknown Split Type.}
Some functions will \emph{change} the split type of a value in an unknown way
upon execution. Ex. 4 in Listing~\ref{listing:string-examples} shows an
example. The \code{filterZeroedRows} function changes the dimensions of its
input, so its output split type is unknown after the call. We represent this in an SA
using a special split type \code{unknown}, which represents a \emph{unique}
split type. Uniqueness prevents pipelining \code{unknown} values with any other
split value: for example, if we tried to pass two \code{unknown} values to
\code{add}, the types would not match, thus preventing pipelining.  However,
generic functions such as \code{scaleMatrix} (Ex. 3), which take a \emph{single}
argument split in any way, can still accept \code{unknown} values. Generics and
\code{unknown} together enable SAs to support operators such as filters, which are
common in data-processing libraries like Pandas.

\subsection{Splitting and Merging with the Splitting API}
\label{subsec:splitting-api}

Annotators bridge the split type abstraction with an implementation using the
splitting API. This API has several roles: it provides
the constructor for constructing parameters, defines how to split data, and
defines how to merge split pieces back into a full result.
Table~\ref{table:sa-api} summarizes these functions.

\paragraph{Constructor.}
The constructor maps values that will appear as function arguments to the split
type's parameters. In our $ArraySplit$ example, the constructor takes a
\code{long} value representing an array's size and returns that size as its
parameter. The constructor should not modify its arguments.

\paragraph{\emph{Split} Function.}
The split function performs the splitting operation using the original function
argument and the split type parameters. It returns a split value representing
the range $[start,end)$ in the function argument. Returning to the MKL
$ArraySplit$ example, the split function would return pointers offset from the
base array pointer. \sn dynamically selects the number of elements in
$[start,end)$ and ensures that $end$ does not surpass the total number of items
in the argument. The \code{Info} function relays information to \sn for this
(\S\ref{sec:runtime}). In our implementation, the split function also takes
additional parameters such as a thread ID and the number of threads.
This allows splits that are not based on integer ranges.

\paragraph{\emph{Merge} Function.}
The associative merge function coalesces split pieces into a single value once \sn finishes
processing them. In our MKL SAs, updates occur in-place, so no merge operation
is needed, but if each function returned a new array instead, the merge
function could concatenate the split arrays into a final result. This function
can also be used to perform operations such as reductions in parallel using
SAs. 

Ex. 5 in Listing~\ref{listing:string-examples} shows an example of a
reduction operator that collapses a matrix into a vector by summing either its
rows or columns. The input matrix is split using $MatrixSplit$ defined
earlier, but the result is a new split type $ReduceSplit \langle axis \rangle$
that represents partial results. In particular, $ReduceSplit$ represents either
reduced row values or column values, depending on \code{axis}: its merge
function uses \code{axis} to reconstruct either a row-vector or a
column-vector.

\input{tables/splitter_api}

\subsection{Conditions to Use SAs}

To summarize, a side-effect-free function $F(a, b, \ldots) \rightarrow c$ can
be annotated with an SA with split types $(A, B, \ldots) \rightarrow C$ if:

\begin{align*}
F(a,b, \ldots) = Merge_C(F(a_1, b_1, \ldots), F(a_2, b_2, \ldots), \ldots)
\end{align*}

\noindent where $Split_A(a) \rightarrow [a_1, a_2, \ldots]$ is the split
function for split type $A$ and $Merge_C(c_1, c_2, \ldots)$ is the associative merge function for
split type $C$. There are no constraints on the number of splits each split
function produces, as long as all split functions produce the same number of
splits for a given function.

\subsection{Summary: How Annotators Use SAs}
To annotate a library, an annotator first decides how to split the library's
core data types. She then defines split types and implements their splitting
APIs. The annotator then writes SAs for side-effect-free functions using the
defined split types. For functions that perform reductions or require custom
merge operations, the annotator implements per-function split types that only
implement the merge function. In our integrations, we required up to three
split types for the core data types per library (e.g., DataFrames in Pandas),
and one split type per reduction function.  We generated most SAs using a
script, since functions with matching function signatures can share the same
SA. We discuss effort of integration in detail in \S\ref{sec:evaluation}.

\subsection{Generality of SAs}

A split type can capture a variety of data formats by splitting inputs and
enabling pipelining and parallelization because annotators define their
splitting behavior. We show in our library integrations in
\S\ref{sec:integration-effort} and in our evaluation in \S\ref{sec:evaluation}
that SAs can split and optimize numerical and scientific workloads that use
arrays, matrices, and tensors, SQL-like workloads that use DataFrames for
operations ranging from projections and selections to groupBys and joins, image
processing workloads, and natural language processing workloads. We note that,
because SAs primarily aim to accelerate data parallel workloads that can be
pipelined (i.e., cases where data movement optimizations in IRs are most
applicable), they will be most impactful over collection-like data.

%% file: tables/splitter_api.tex
\begin{table}[]
\small
\begin{tabular}{l}
\toprule
  \textbf{Splitting API Summary (\S\ref{subsec:splitting-api})}                                                                    \\ \midrule
\code{NameConstructor(A0,...An) => Parameters}                                    \\ \midrule
\code{Split(D arg, int start, int end, Parameters) => D}           \\ \midrule
\code{Merge(Vector<D>, Parameters) => D} \\ \midrule
\code{Info(D arg, Parameters) => RuntimeInfo}           \\ \bottomrule
\end{tabular}
\caption{The API annotators implement for a split type $Name
  \langle Parameters \rangle$. The argument has a data type \code{D}.}
\label{table:sa-api}
\vspace{-1em}
\end{table}

%% file: client_library.tex
\section{The \sn Client Libraries}
\label{sec:client-library}

\sn relies on a lazily evaluated dataflow graph to enable cross-function data
movement optimizations. Nodes in the dataflow graph represent calls to
annotated functions and their arguments, and edges represent data passed
between functions. Constructing such a graph without library modifications
(e.g., by using an API as in prior work~\cite{weld-arxiv,tf-ops}) is
challenging because applications will contain a mix of annotated function calls
and arbitrary code that may access lazy values. The \libsn client library is
responsible for capturing a graph and determining when to evaluate it. The
library has a small interface: \emph{register(function, args)} registers a
function and its arguments with the dataflow graph, and \emph{evaluate()}
evaluates the dataflow graph (\S\ref{sec:runtime} describes the runtime) when
arbitrary code accesses lazy values. Since the \libsn design is coupled with
the annotated library's language, we discuss its design in two languages: C++
and Python.

 \begin{figure}[t!]
  \centering
  \includegraphics[width=\columnwidth]{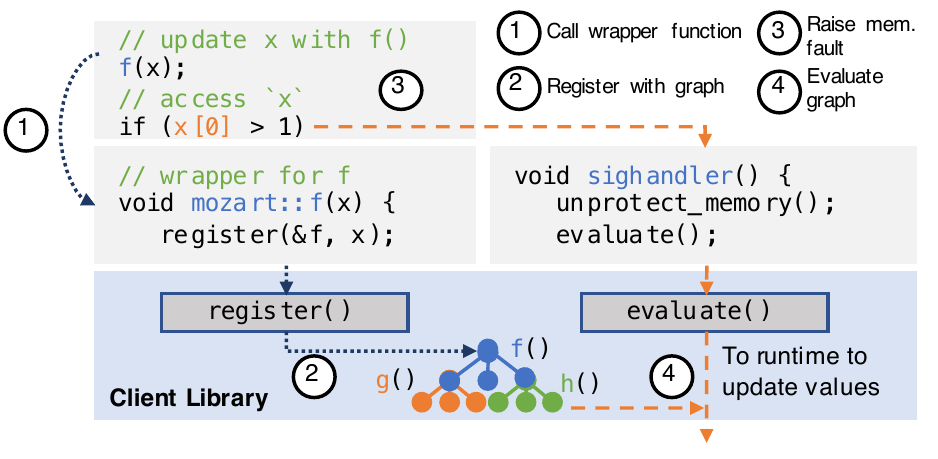}
  \caption{Overview of the C++ client library.}
  \label{fig:clc}
\end{figure}

\subsection{C++ Client Library}
\label{subsec:c-client-library}

Our C++ client library uses code generation and OS-level memory
protection to build a dataflow graph and to determine when to evaluate it.
Figure~\ref{fig:clc} outlines its design.

\paragraph{Writing Annotations.} An annotator registers split types, the
splitting API, and SAs over C++ functions by using a command line tool we have built
called \code{annotate}. This tool takes these definitions as input and
generates namespaced C++ \emph{wrapper functions} around each annotated library
function. These wrapper functions are packaged with the splitting API and a
lookup table that maps functions to their SAs in a shared library. The
application writer links this wrapped library and calls the wrapper functions
instead of the original library functions as always. This generally requires a
namespace import and no other code changes--we note one exception below.

\paragraph{Capturing a Graph.}
The wrapper functions are responsible for registering tasks in the dataflow
graph. When an application calls a wrapper, its function arguments are copied
into a buffer, and the \libsn \emph{register} API adds the function and its
argument buffer as a node in the dataflow graph. The wrapper knows which
function arguments are mutable, based on which arguments in the SA were marked
\code{mut} (the SA is retrieved using the lookup table). This allows \libsn to
add the correct data-dependency edges between calls.

\paragraph{Determining Evaluation Points.}
We evaluate the dataflow graph upon access to lazy values. There are two
cases: \one the accessed value was returned by an annotated function, and \two
the accessed value was allocated outside of the dataflow graph but mutated by
an annotated function.

\lstset{language=CustomC} To handle the first case, if the library function
returns a value, its wrapper instead returns a type called \code{Future<T>}.
For types where \code{T} is a pointer, \code{Future<T>} is a pointer with an
overloaded dereference operator that first calls \emph{evaluate} to evaluate
the dataflow graph. For non-pointer values, the value can be accessed
explicitly with a \code{get()} method, which forces the execution. We also
override the copy constructor of this type to track aliases, so copies of a
lazy value can be updated upon evaluation. Wrapper functions can accept both
\code{Future<T>} values and \code{T} values, so \code{Future} values may be
pipelined. Usage of \code{Future<T>} is the main code change applications must
make when using SAs in C++.

We handle the second case (shown in Figure~\ref{fig:clc} in the access to
\code{x[0]}) by using memory protection to intercept reads of lazy values.
Applications must use our drop-in \code{malloc} and \code{free} functions for
memory accessed in the dataflow graph (e.g., via \code{LD_PRELOAD}). Our
version of \code{malloc} uses the \code{mmap} call to allocate memory with
\code{PROT_NONE} permissions, which raises a protection violation when read or
written. \libsn registers a signal handler to catch the violation, unprotects
memory, and evaluates the dataflow graph registered so far. When calling a
wrapper function for the first time after evaluation, \libsn re-protects all
allocated memory to re-enable capturing memory accesses. This 
technique has been used in other systems
successfully~\cite{kristensen2014bohrium,kotthaus2015dynamic} to inject
laziness.

\subsection{Python Client Library}
\label{subsec:implementation:python}

\lstset{language=AnnotationC}

\paragraph{Writing Annotations.}
Developers provide SAs by using Python function decorators. In Python, split types for
positional arguments are required, and split types for keyword arguments
default to ``\code{_}'' (but can be overridden).

\paragraph{Capturing a Graph.} \libsn constructs the dataflow graph using the
same function decorator used to provide the SA. The decorator wraps the
original Python function into one that records the function with the graph
using \emph{register()}. The wrapper function then returns a placeholder
\code{Future} object.

\paragraph{Determining Evaluation Points.}
Upon accessing a \code{Future} object, \libsn evaluates the task
graph. In Python, we can detect when an object is accessed by overriding its
builtin methods (\eg \code{__getattribute__} to get object attributes).
After executing the task graph,
the \code{Future} object forwards calls to these methods to the evaluated
cached value. To intercept accesses to variables that are mutated by annotated
functions (based on \code{mut}), when \libsn registers a mutable object, it
overrides the object's \code{__getattribute__} to again force evaluation when
the object's fields are accessed. The  original \code{__getattribute__} is
reinstated upon evaluation.

%% file: runtime.tex
\section{The \sn Runtime}
\label{sec:runtime}

\eat {
\begin{figure}[t!]
  \centering
  \includegraphics[width=0.48\textwidth]{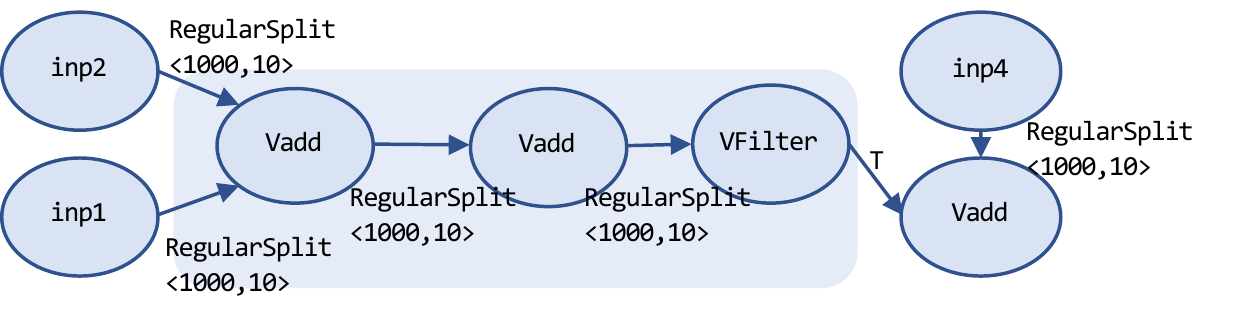}
  \caption{Converting a dataflow graph into stages.}
  \label{fig:dag}
\end{figure}
}

\sn is a parallel runtime that executes functions annotated with SAs. \sn takes
the dataflow graph generated by the client library and converts it into an
\emph{execution plan}. Specifically, \sn converts the dataflow graph into a
series of \emph{stages}, where each stage splits its inputs, pipelines the
split inputs through library functions, and then merges the outputs. The SAs
dictate the stage boundaries. \sn then executes each stage over \emph{batches}
in parallel, where each batch represent one set of split inputs.

\subsection{Converting a Dataflow Graph to Stages}
\label{subsec:type-inference}

\lstset{language=AnnotationC}

Recall that each node in the dataflow graph is an annotated function call, and
each edge from function $f_1$ to $f_2$ represents a data dependency, i.e., a
value mutated or returned by $f_1$ and read by $f_2$. \sn converts this graph
into stages. The functions $f_1$ and $f_2$ are in the same stage if, for every
edge between them, the source value and destination value have the same split
type. If \emph{any} split types between $f_1$ and $f_2$ do not match, split
data returned by $f_1$ must be merged, and a new stage starts with $f_2$. \sn
traverses the graph and checks types to construct stages.

To check split types between a source and destination argument, \sn first
checks that the split types have the same name. If the names are equal, \sn
uses the function arguments captured as part of the dataflow graph to
initialize the split types' parameters. If the parameters also match, the
source and destination have the same split type. If either split type is a
generic, \sn uses type inference~\cite{aiken1993type,pierce2000local} to
determine its type by pushing known types along the edges of the graph to set
generics. If a split type cannot be inferred (e.g., because all functions use
generics), \sn falls back to a default for the data type: in our
implementation, annotators provide a default split type constructor per data
type.

This step produces an execution plan, where each stage contains an ordered list
of functions to pipeline. The inputs to each stage are split based on the
inputs' split types, and the outputs are merged before being passed to the next
stage.

\subsection{Execution Engine}
\label{subsec:parallelism}

After constructing stages, \sn executes each stage in sequence by \one choosing
a batch size, \two splitting and executing each function, and
\three merging partial results. 

\paragraph{Step 1: Discovering Runtime Parameters.} \sn sets the number of
elements in each batch and the number of elements processed per thread as
runtime parameters. Since the goal of pipelining is to reduce data movement, we
use a simple heuristic for the batch size: each batch should contain roughly
\emph{sizeof(L2 cache)} bytes.  To determine the batch size, \sn calls each
input's \code{Info} function (\S\ref{sec:annotations}), which fills a struct
called \code{RuntimeInfo}. This struct specifies the number of total elements
that will be produced for the input (e.g., number of elements in an array or
number of rows in a matrix), and the size of each element in bytes. The batch
size is then set to $\frac{C \times L2CacheSize}{\sum{sizeof(element)}}$ (where
$C$ is a fixed constant). We found that this value works well empirically (see
\S\ref{sec:evaluation}) when pipelines allocate intermediate split values too,
since these values are still small enough to fit in the larger shared
last-level-cache.

Workers partition elements equally among themselves. The user configures the
number of workers.  \sn checks to ensure that each split produces the same
total number of elements. We opted for static parallelism rather than dynamic
parallelism (e.g., via work-stealing) because it is simpler to schedule and we
found that it leads to similar results for most workloads: however, dynamic
work-stealing schedulers such as Cilk~\cite{cilk} are also compatible with \sn.

\paragraph{Step 2: Executing Functions.} After setting runtime parameters, \sn
spawns worker threads and communicates to each thread the range of elements it
processes.  Each worker allocates thread-local temporary buffers for the split
values and enters the main driver loop. The worker's driver loop calls the
\emph{Split} function for each input and writes the result into the temporary
buffers. If \emph{Split} returns \code{NULL}, the driver loop exits. For
arguments with the missing ``\code{_}'' split type, the original input value is
copied (usually, this is just a pointer-copy) rather than split. The
\emph{start} and \emph{end} arguments of the \emph{Split} function are set
based on the batch size and the thread's range.

To execute the function pipeline per thread, \sn tracks which temporary buffers should be
fed to each function as arguments by using a mapping from unique
argument IDs to buffers. The execution plan represents function calls using
these argument IDs (e.g., a call $f_1(a0,a1,a2) \rightarrow a3$ will pass the
buffers for $a0$, $a1$, and $a2$ as arguments and store the result in the
buffer for $a3$). After each batch, these buffers are moved to a list of
partial results, and \sn starts the next batch.

\paragraph{Step 3: Merging Values.} Once the driving loop exits, each worker merges each list of
temporary buffers via the split type merge function (the stage
tracks the split type of each result), and then returns the merged result. Once all
workers return their partial results, \sn calls the merge function again on the
main thread to compute the final merged results.

%% file: implementation.tex
\section{Implementation}
\label{sec:implementation}

Our C++ version of \libsn and \sn is implemented in roughly 3000 lines of
Rust. This includes the parser for the SAs, the \code{annotate} tool for
generating header files containing the wrapper functions, the client library
(including memory protection), planner, and parallel runtime.  We use Rust's
threading library for parallelism. We make heavy use of the \code{unsafe}
features of Rust to call into C/C++ functions. Memory allocated for splits is
freed by the corresponding mergers. \sn manages and frees temporary memory.

The Python implementation of these components is in around 1500 lines of
Python. The SAs themselves use Python's function decorators, and split types
are implemented as abstract classes with splitter and merger methods. We use
process-based parallelism to circumvent Python's global interpreter lock. For
native Python data, we only need to serialize data when communicating results
back to the main thread. We leverage copy-on-write \emph{fork()} semantics when
starting workers, which also means that ``\code{_}'' values need not be cloned.

%% file: integration.tex
\section{Library Integrations}
\label{sec:integration-effort}

We evaluate SAs by integrating them with five popular data processing
libraries: NumPy~\cite{numpy}, Pandas~\cite{pandas} spaCy~\cite{spacy}, Intel
MKL~\cite{mkl} and ImageMagick~\cite{imagemagick}.
Table~\ref{table:integration-effort} in \S\ref{sec:evaluation} summarizes
effort, and we discuss integration details below.

\paragraph{NumPy.} NumPy is a popular Python numerical processing library, with
core operators implemented in C. The core data type in the library is the
\code{ndarray}, which represents an N-dimensional tensor. We implemented a
single split type for \code{ndarray}, whose splitting behavior depends on its
shape and the axis a function iterates over (the split type's constructor maps
\code{ndarray} arguments to its shape). We added SAs over all tensor unary, 
binary, and associative reduction operators. We implemented split types for
each reduction operator to merge the partial results: these only required merge
functions.

\lstset{language=CustomC}

\paragraph{Pandas.} Our Pandas integration implements split types over
DataFrames and Series by splitting by row. We also added a $GroupSplit$ split type for
\code{GroupedDataFrame}, which is used for groupBy operations. Aggregation
functions that accept this split type group chunks of a DataFrame, create
partial aggregations, and then re-group and re-aggregate the partial
aggregations in the merger. We only support commutative aggregation functions.
We support most unary and binary Series operators, filters, predicate
masks, and joins: joins split one table and broadcast the other.
\lstset{language=AnnotationC}
Filters and joins return the \code{unknown} split type, and most 
functions accept generics.
\lstset{language=CustomC}

\paragraph{spaCy.} SpaCy is a Python natural language processing library
with operators in Cython. We added a split type that uses spaCy's builtin minibatch
tokenizer to split a corpus of text. This allows any function (including
user-defined ones) that accepts text and internally uses spaCy functions to be
parallelized and pipelined via a Python function decorator.

\paragraph{Intel MKL.} Intel MKL~\cite{mkl} is an optimized closed-source
numerical computation library used as the basis for other computational
frameworks~\cite{pytorch,abadi2016tensorflow,eigen,kristensen2014bohrium,numpy,numba}. To integrate
SAs, we defined three split types: one for
matrices (with rows, columns, and order as parameters), one for arrays (with
length as a parameter), and one for the size argument. Since MKL operates over
inputs in place, we did not need to implement merger functions. We annotated
all functions in the vector math header, the saxpy (L1 BLAS) header, and the
matrix-vector (L2 BLAS) headers.

\paragraph{ImageMagick.} ImageMagick~\cite{imagemagick} is a C image processing
library that contains an API where images are loaded and processed using an
opaque handle called \code{MagickWand}. We implemented a split type for the
\code{MagickWand} type, where the split function uses a crop function to clone
and return a subset of the original image. ImageMagick also contains an API for
appending several images together by stacking them---our split function thus
returns entire rows of an image, and this API is used in the merger to
reconstruct the final result.

\subsection{Experiences with Integration}

\paragraph{Unsupported Functions.}
We found that there were a handful of functions in each library that we could
not annotate. For example, in the ImageMagick library, the \code{Blur} function
contains a boundary condition where the edges of an image are processed
differently from the rest of the image. SAs' split/merge paradigm would produce
incorrect results here, because this special handling would occur on each split
rather than on just the edges of the full image. Currently, annotators must
manually determine whether each function is safe to annotate (similar to other
annotation-based systems such as OpenMP). We found that this was
straightforward in most cases by reading the function documentation, but tools
that could formally prove an SA's compatibility with a function would be
helpful. We leave this to future work. We did not find any functions that internally held locks or were not
callable from multiple threads in the libraries we annotated.

\paragraph{Debugging and Testing.}
Since annotators must manually enforce the soundness of an SA, we built some
mechanisms to aid in debugging and testing them. The \code{annotate} tool, for
example, will ensure that a split type is always associated with the same
concrete type. The runtimes for both C and Python include a ``pedantic mode''
for debugging that can be configured to panic if a function receives splits
with differing numbers of elements, receives no elements, or receives \code{NULL}
data. The runtime can also be configured to log each function call on each
split piece, and standard tools such as Valgrind or GDB are still available.
Anecdotally, the logs and pedantic mode made debugging invalid split/merge
functions and errors in SAs unchallenging. We also fuzz tested our annotated
functions to stress their correctness. 

%% file: evaluation.tex
\section{Evaluation}
\label{sec:evaluation}

In evaluating \sas, we seek to answer several questions:
\one Can SAs accelerate end-to-end workloads that use existing unmodified libraries?
\two Can SAs match or outperform compiler-based techniques that optimize and JIT machine code?
\three Where do performance benefits come from, and which classes of workloads do SAs help the most?

\input{tables/workloads}

\paragraph{Experimental Setup.} 
We evaluate workloads that use the five libraries in
\S\ref{sec:integration-effort}: NumPy v1.16.2, Pandas v0.22.0, spaCy v2.1.3,
Intel MKL 2018 and ImageMagick v7.0.8. We ran
all experiments on an Amazon EC2 \code{m4.10xlarge} instance with Intel Xeon
E5-2676 v3 CPUs (40 logical cores) and 160GB of RAM, running Ubuntu 18.04
(Linux 4.4.0). Unless otherwise noted, results average over five runs.

\subsection{Workloads}
\label{subsec:workloads}
We evaluate the end-to-end performance benefits of SAs using \sn
on a suite of 15 data analytics workloads (of which four are repeated in NumPy
and Intel MKL). Eight of the benchmarks are taken from the Weld
evaluation~\cite{palkar2018evaluating}, which obtained them from popular GitHub
repositories, Kaggle competitions, and online tutorials. We also evaluate an
additional numerical analysis workload (Shallow Water) over matrix operations,
taken from the Bohrium paper~\cite{kristensen2014bohrium} (Bohrium is an optimizing NumPy
compiler that we compare against here). Finally, we evaluate two
open-source image processing workloads~\cite{instagram-filters} that use
ImageMagick, and a part-of-speech tagging workload~\cite{spacy} that uses spaCy.

\subsection{End-to-end Performance Results}
\label{subsec:eval:e2e}

\begin{figure*}[t!]
  \centering
  \includegraphics[width=\textwidth]{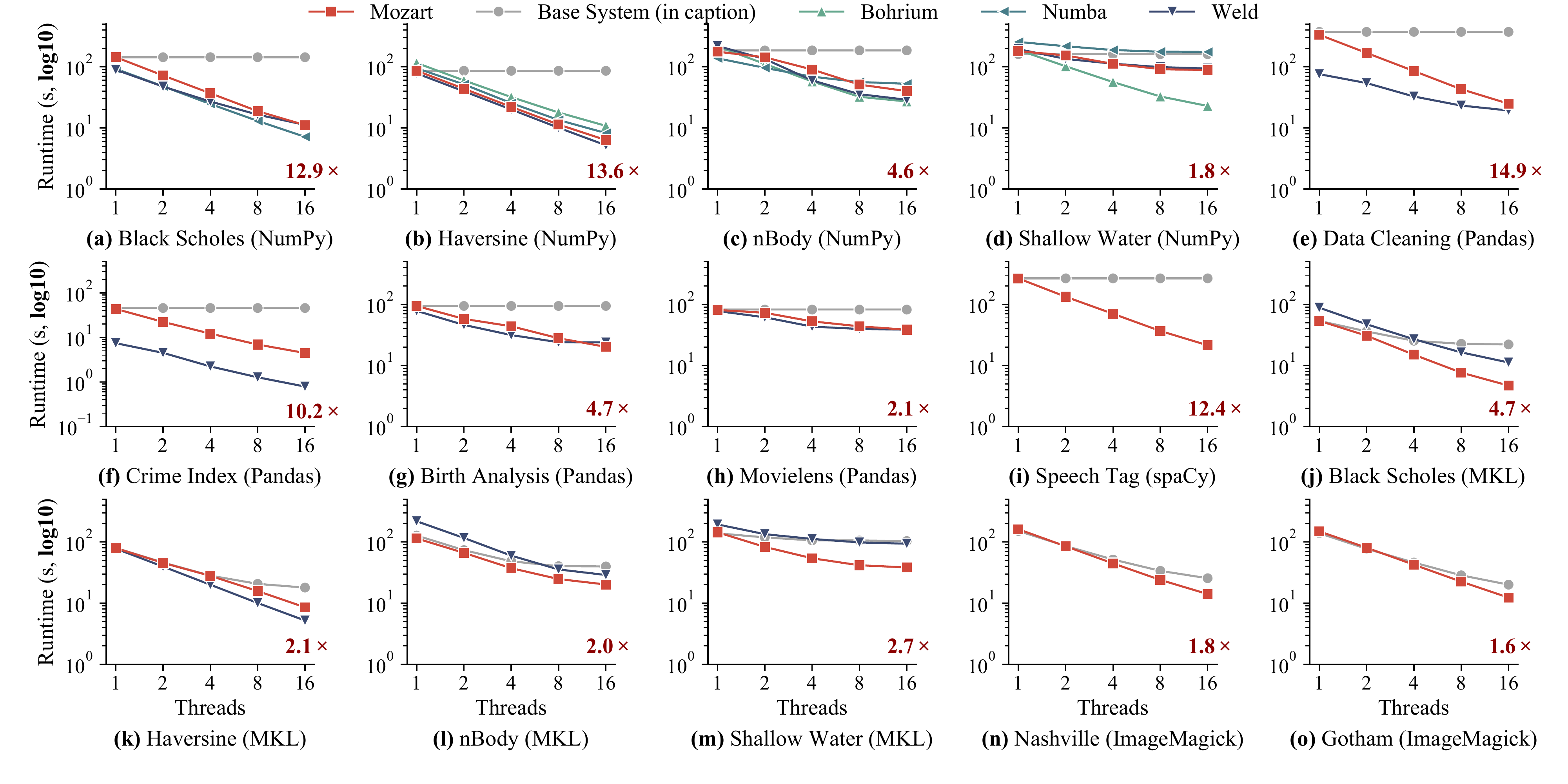}
  \caption{End-to-end performance on 15 benchmarks compared against a base system (in caption, e.g., NumPy) and several optimizing compilers that require rewriting libraries. We show results on 1--16 threads. Each plot displays the speedup (in red) that \sn enables on 16 threads over the base system.} 
  \label{fig:e2e}
\end{figure*}

Figure~\ref{fig:e2e} shows \sn's end-to-end performance on our 15 benchmarks on
1--16 threads. Each benchmark compares \sn vs. a base system without SAs (e.g.,
NumPy or MKL without SAs). We also compare against optimizing
compiler-based approaches that enable parallelism and data movement
optimization without changing the library's API, but require
\emph{re-implementing functions} under-the-hood.

\paragraph{Summary of all Results.} On 16 threads, \sn provides speedups of up
to 14.9$\times$ over libraries that are single-threaded, and speedups of up to
4.7$\times$ over libraries that are already parallelized due to its data
movement optimizations. Across the 15 workloads, \sn is within 1.2$\times$ of
all compilers we tested in six workloads, and outperforms all the
compilers we tested in four workloads. Compilers outperform \sn by over
1.5$\times$ in two workloads. Compilers have the greatest edge over \sn in
workloads that contain many operators implemented in interpreted Python, since
they naturally benefit more from compilation to native code than they do from
data movement optimizations. We discuss workloads below.

\paragraph{NumPy Numerical Analysis (Figures~\ref{fig:e2e}a-d).}
We evaluate the NumPy workloads against un-annotated NumPy and three
Python JIT compilers: Bohrium~\cite{kristensen2014bohrium}, Numba~\cite{numba}, and Weld.
Each compiler requires function rewrites under-the-hood.
Figures~\ref{fig:e2e}a-b show the performance of Black Scholes and Haversine,
which apply vector math operators on NumPy arrays. All systems enable
near-linear scalability since all operators can be pipelined and parallelized.
Overall, \sn enables speedups of up to 13.6$\times$. 

The nBody and Shallow Water workloads operate over tensors and matrices, and
contain operators that cannot be pipelined. For example, Shallow Water performs
several row-wise matrix operations and then aggregates along columns to compute
partial derivatives. \sn captures these boundaries using split types and still
pipelines the other operators in these workloads.  Figures~\ref{fig:e2e}c-d
show that \sn enables up to 4.6$\times$ speedups on 16 threads. Bohrium
outperforms other systems in the Shallow Water benchmark because it captures
indexing operations that \sn cannot split and that the other compilers could
not parallelize (Bohrium converts the indexing operation into its
IR, whereas \sn treats it as a function call over a single
element that cannot be split).

\paragraph{Data Science with Pandas and NumPy (Figures~\ref{fig:e2e}e-h).} We
compare the Pandas workloads against Weld: the other compilers did not
accelerate these workloads. The Data Cleaning and Crime Index workloads use
Pandas and NumPy to filter and aggregate values in a Pandas DataFrame.
Figures~\ref{fig:e2e}e-f show the results. \sn parallelizes and pipelines both
of these workloads and achieves an up to 14.9$\times$ speedup over the native
libraries. However, Weld outperforms \sn by up to 5.85$\times$ even on 16
threads, because both contain operators that use interpreted Python code which
Weld compiles.

Figures~\ref{fig:e2e}g-h shows the results for the Birth Analysis and MovieLens
workloads. These workloads are primarily bottlenecked on grouping and joining
operations implemented in C. In Birth Analysis, \sn accelerates groupBy
aggregations by splitting grouped DataFrames and parallelizing (there are no
pipelined operators), leading to a 4.7$\times$ speedup. In MovieLens, we
pipeline and parallelize two joins and parallelize a grouping
aggregation, leading to a 2.1$\times$ speedup. In both, \sn outperforms Weld.
Weld's parallel grouping implementation bottlenecked on memory allocations
around 8 threads in Birth Analysis. In MovieLens, speedups were hindered due to
serialization overhead (Weld marshals strings before processing them, and \sn
sends large join results via IPC), but the Weld serialization could not
be parallelized.

\paragraph{Speech Tagging with spaCy (Figures~\ref{fig:e2e}i).}
Figure~\ref{fig:e2e}i shows the performance of the speech tagging workload with
and without \sn. This workload operates over a corpus of text from the
IMDb sentiment dataset~\cite{imdb-database}. It tags each word with a part of speech and
normalizes sentences using a preloaded model. \sn enables 12$\times$
speedups via parallelization.  Unfortunately, no compilers supported spaCy.

\paragraph{Numerical Workloads with MKL (Figures~\ref{fig:e2e}j-m).} We
evaluate \sn with MKL using the same numerical workloads from NumPy. Unlike
NumPy, MKL already parallelizes its operators, so the speedups over it come
from optimizing data movement.  Figures~\ref{fig:e2e}j-m show that \sn improves
performance by up to 4.7$\times$ on 16 threads, even though MKL also
parallelizes functions. \sn outperforms Weld on three
workloads here, because MKL vectorizes and loop-blocks matrix operators in
cases where Weld's compiler does not.

\paragraph{Image workloads in ImageMagick (Figures~\ref{fig:e2e}n-o).}
Figure~\ref{fig:e2e} shows the results on our two ImageMagick workloads.  Like
MKL, ImageMagick also already parallelizes functions, but \sn accelerates them by
pipelining across operators. \sn outperforms base ImageMagick by up to
1.8$\times$. End-to-end speedups were limited despite pipelining because
splits and merges allocate and copy memory excessively. \sn
sped up just the computation by up to 3.4$\times$ on 16 threads.

\subsection{Effort of Integration}
\label{subsec:integration}

In contrast to compilers, \sn only requires annotators to add SAs and implement
the splitting API to achieve performance gains. To quantify this effort, we
compared the lines of code required to support our benchmarks with \sn vs.
Weld.  Table~\ref{table:integration-effort} shows the results. Our Weld results
only count code for operators that we also support, and only counts integration
code.  We do not count code for the Weld compiler itself (which itself is over
25K LoC and implements a full compilation backend based on LLVM). Similarly, we
only count code that an annotator adds for \sn, and do not count the runtime
code.

Overall, for our benchmarks, SAs required up to 17$\times$ less code to enable similar
performance to Weld in many cases. Weld required at least tens of lines of IR code per
operator, a C++ extension to marshal Python data, and usage of its runtime API
to build a dataflow graph. Anecdotally, through communication with the
Pandas-on-Weld authors, we found that just supporting the Birth Analysis
workload, even after having implementing the core integration, was a multi-week
effort that required several extensions to the Weld compiler itself and also
required extensive low-level performance tuning. In contrast, we integrated SAs
with the same Pandas functions in roughly half a day, and the splitting API was
implemented using existing Pandas functions with fewer than 20 LoC each.

\subsection{Importance of Pipelining}
\label{subsec:pipelining}

\input{tables/integration-effort}
\input{tables/perf-data}

The main optimization \sn applies beyond parallelizing split data is pipelining
it across functions to reduce data movement. To show its importance,
Table~\ref{table:pipelining} compares three versions of the Black Scholes and
Haversine workloads on 16 threads: un-annotated MKL, \sn with pipelining, and
\sn without pipelining (i.e. \sn parallelizes functions on behalf of MKL).  We
also used the Linux \code{perf} utility to sample the hardware performance
counters for the three variations of the workload. \sn without pipelining does
not result in a speedup over parallel MKL.  In addition, the most notable
difference with pipelining is in the last level cache (LLC) miss rate: the miss
rate decreases by a factor of 2$\times$, confirming that pipelining does indeed
generate less traffic to main memory and reduce data movement. This in turn
leads to better overall performance on multiple threads. We found no other
notable differences in the other reported counters. We saw similar results for
the other MKL workloads as well.

\subsection{System Overheads}

\begin{figure}[t]
  \centering
  \includegraphics[width=0.48\textwidth]{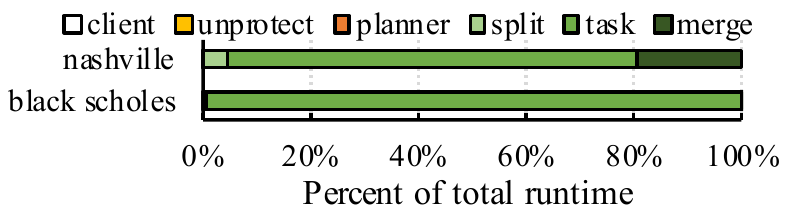}
  \caption{Breakdown of total running time in the Nashville and Black Scholes workloads. Across all workloads, we observed 0.5\% overhead from other components.}
  \label{fig:breakdown}
\end{figure}

To measure the system overheads that \sn imposes, Figure~\ref{fig:breakdown}
shows the breakdown in running time for the Black Scholes and Nashville
workloads on 16 threads. We report the client library time for registering
tasks, the memory protection time for unprotecting pages during execution,
planning, splitting, task execution, and merging. The Nashville workload had
the highest relative split and merge times, since both the splitters and
mergers allocate memory and copy data. Across all workloads, the execution time
dominates the total running time, and the client library, memory protection,
and planner account for less than 0.5\% of the running time.

Most of the overhead is attributed to handling memory protection. In our setup,
unprotecting each gigabyte of data took roughly 3.5ms, indicating that this
overhead could be significant on task graphs that perform little computation.
One mechanism for reducing the overhead of memory protection is the recent
\code{pkeys}~\cite{pkeys} set of system calls, which allows for O(1) memory
permission changes by associating pages with a registered protection key. After
tagging memory pages with a key, the cost of changing their permissions is a
register write, so the time to unprotect or protect \emph{all} memory allocated
with \sn becomes negligible (tens of microseconds to unprotect 1GB in a
microbenchmark we ran).

\subsection{Effect of Batch Size}
\lstset{language=CustomC}

\begin{figure}[t!]
  \centering
  \begin{subfigure}[b]{0.48\columnwidth}
    \centering
    \includegraphics[width=35mm]{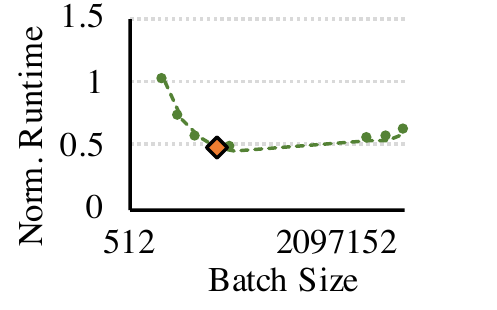}
    \caption{Black Scholes.}
    \label{fig:bs-batch}
  \end{subfigure}
  \begin{subfigure}[b]{0.48\columnwidth}
    \centering
    \includegraphics[width=35mm]{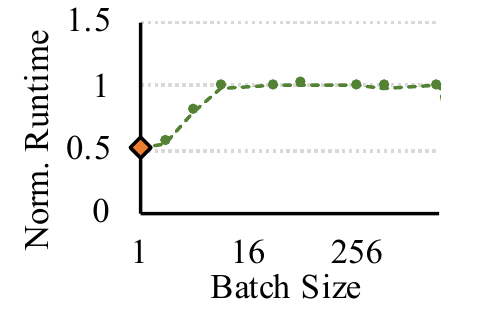}
    \caption{nBody.}
    \label{fig:nbody-batch}
  \end{subfigure}
  \caption{
  Effect of batch size on two workloads. \sn
  selects a batch size near the optimal using L2 cache size.
  }
  \label{fig:batch-size}
\end{figure}

We evaluate the effect of batch size by varying it on the Black Scholes and
nBody workloads and measuring end-to-end performance for each parameter.  We
benchmarked these two workloads because they contain ``elements'' of different
sizes: Black Scholes treats each \code{double} as a single element, while the
matrices' split types in nBody treat rows of a matrix (256KB in size each) as a
single element. Figure~\ref{fig:batch-size} shows the results. The marked point
shows the batch size selected by \sn using the strategy described in
\S\ref{sec:runtime}. The plots show that batch size can have a significant
impact on overall running time (too low imposes too much overhead, and too high
obviates the benefits of pipelining), and that \sn's heuristic scheme selects a
reasonable batch size. Across all the workloads we benchmarked, \sn chooses a
batch size within 10\% of the best batch size we observed in a parameter sweep.

\subsection{Compute- vs. Memory-Boundedness} 
\begin{figure}[t!]
  \centering
  \begin{subfigure}[b]{0.48\columnwidth}
    \centering
    \includegraphics[width=35mm]{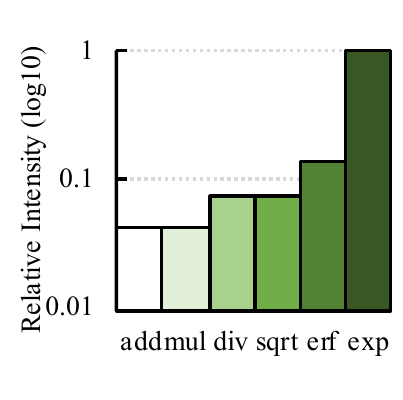}
    \caption{Relative Intensity}
    \label{fig:intensity-relative}
  \end{subfigure}
  \begin{subfigure}[b]{0.48\columnwidth}
    \centering
    \includegraphics[width=35mm]{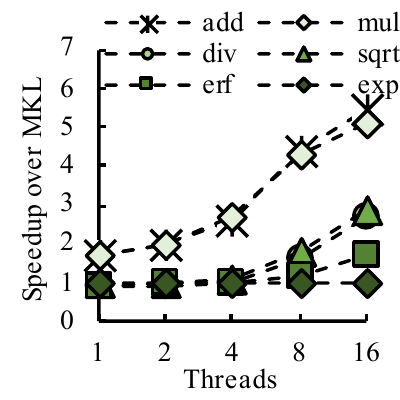}
    \caption{Speedup over no SAs}
    \label{fig:intensity-speedup}
  \end{subfigure}
  \caption{Impact of compute-intensiveness in \sn.}
  \label{fig:intensity-overview}
\end{figure}

To study when \sn's data movement
optimizations are most impactful, we measured the intensity (defined as
\emph{cycles spent per byte of data}) of several MKL vector math functions by
calling them in a tight loop on an array that fits entirely in the L2 cache. We
benchmarked the following operations, in order of increasing intensity: add,
mul, sqrt, div, erf, and exp.  Figure~\ref{fig:intensity-relative} shows the
relative intensities of each function (i.e., \code{vdExp} spends roughly
7$\times$ more cycles per byte of data than \code{vdErf}).  We then ran each
math function 10 times on a large 8GB array, with and without \sn.
Figure~\ref{fig:intensity-speedup} shows the \emph{speedup} of \sn over
un-annotated MKL on 1--16 threads. \sn has the largest impact on
memory-intensive workloads that spend few cycles per byte, and shows increasing
speedups as increasing amounts of parallelism starve the available memory bandwidth.

%% file: tables/workloads.tex
\begin{table}[t!]
\small
\setlength\tabcolsep{3pt} 
\centering
  \begin{tabular}{@{}p{0.7in}p{0.68in}p{1.75in}@{}}
\toprule
\multicolumn{1}{c}{\textbf{Workload}}   & \textbf{Libraries}                                         & \multicolumn{1}{c}{\textbf{Description (\# Operators)}}                                                                                       \\ \midrule
    Black \hspace{4em} Scholes                           & NumPy, \hspace{4em} MKL        & Computes the Black Scholes~\cite{blackscholes} formula over a set of vectors. (32)                                                            \\ \midrule
    Haversine                               & NumPy, \hspace{4em} MKL        & Computes Haversine Dist.~\cite{haversine} from a set of GPS coordinates to a fixed point. (18)                                                \\ \midrule
    nBody                                  & NumPy, \hspace{4em} MKL        & Uses Newtonian force equations to determine the position/velocity of stars over time. (38)                        \\ \midrule
    Shallow \hspace{4em} Water                           & NumPy, \hspace{4em} MKL        & Estimates the partial differential equations used to model the flow of a disturbed fluid~\cite{shallowwater}. (32)                            \\ \midrule
    Data \hspace{4em} Cleaning~\cite{datacleaning}       & Pandas            & Cleans a DataFrame of 311 requests~\cite{311requests} by replacing \code{NULL}, broken, or missing values with \code{NaN}. (8)                \\ \midrule
    Crime Index                             & Pandas, \hspace{4em} NumPy     & Computes an average ``crime index'' score, given per-city population and crime information. (16)                                              \\ \midrule
    Birth \hspace{4em} Analysis~\cite{birthanalysis}     & Pandas, NumPy     & Given a dataset of number of births by name/year, computes fraction of names starting with ``Lesl'' grouped by gender and year-of-birth. (12) \\ \midrule
    MovieLens                               & Pandas, \hspace{4em} NumPy     & Joins tables from the MovieLens dataset~\cite{movielens} to find movies that are most divisive by gender. (18)                   \\ \midrule
    Speech \hspace{4em} Tag                 & spaCy     &                    Tags parts of speech and extracts features from a corpus of text. (8)                   \\ \midrule
Nashville                               & ImageMagick       & Image pipeline~\cite{instagram-filters-nash} that applies color masks, gamma correction, and HSV modulation.  (31)                \\ \midrule
 Gotham                                 & ImageMagick       & Image pipeline~\cite{instagram-filters-gotham} that applies color masks, saturation/contrast adjustment, and modulation. (15)       \\ \midrule
\end{tabular}
\caption{Workloads used in our evaluation. Descriptions for workloads from Weld
  are taken from~\cite{palkar2018evaluating}. The number in parentheses shows the number of
  library API calls.}
\label{table:workloads}
\end{table}

%% file: tables/integration-effort.tex
\begin{table}[]
\small
\setlength\tabcolsep{2pt} 
\begin{tabular}{@{}lcccc|ccc@{}}
\toprule
\multicolumn{1}{c}{\textbf{}} & \multicolumn{1}{l}{\textbf{}}                                                          & \multicolumn{3}{c}{\textbf{LoC for SAs}}            & \multicolumn{3}{c}{\textbf{LoC for Weld}}             \\ \cmidrule(l){3-8} 
Library             & \#Funcs & SAs & Split. API & Total & Weld IR  & Glue & Total \\ \midrule
NumPy                         & 84                                                                                     & 47  & 37  & \textbf{84}    & 321  & 73    & \textbf{394}                       \\
Pandas                        & 15                                                                                     & 72  & 49  & \textbf{121}   & 1663 & 413   & \textbf{2076}                      \\
spaCy                         & 3                                                                                      & 8   & 12  & \textbf{20}    &      &       & \textbf{}                          \\
MKL                           & 81                                                                                     & 74  & 90  & \textbf{155}   &      &       & \textbf{}                          \\
ImageMagick                   & 15                                                                                     & 49  & 63  & \textbf{112}   &      &       & \textbf{}                          \\ \bottomrule
\end{tabular}
  \caption{Integration effort for using \sn. Numbers show the total lines of code per library. \sn requires up to 15$\times$ fewer LoC to support the same operators as Weld.}
\label{table:integration-effort}
\end{table}

%% file: tables/perf-data.tex
\begin{table}[]
\small
\setlength\tabcolsep{2pt} 
\begin{tabular}{@{}lcccccc@{}}
\toprule
\textbf{}                                                                  & \multicolumn{3}{c}{Black Scholes}                                                                                                                                                                                  & \multicolumn{3}{c}{Haversine}                                                                                                                                                                 \\ \cmidrule(l){2-7} 
\textbf{}                                                                  & MKL                                                        & \begin{tabular}[c]{@{}c@{}}\sn\\ (-pipe)\end{tabular}      & \multicolumn{1}{c|}{\textbf{\sn}}                                                        & MKL                                                        & \begin{tabular}[c]{@{}c@{}}\sn\\ (-pipe)\end{tabular}      & \textbf{\sn}                                                        \\ \midrule
  \textbf{\begin{tabular}[c]{@{}l@{}}Normalized\\Runtime\end{tabular}} &    1.00   & 1.01     & \textbf{0.21}     & 1.00     & 0.97     &  \textbf{0.48}    \\ \midrule
\textbf{\begin{tabular}[c]{@{}l@{}}LLC Miss\\ (avg/stddev)\end{tabular}}   & \begin{tabular}[c]{@{}c@{}}39.86\%\\ (2.26\%)\end{tabular} & \begin{tabular}[c]{@{}c@{}}42.96\%\\ (1.56\%)\end{tabular} & \multicolumn{1}{c|}{\textbf{\begin{tabular}[c]{@{}c@{}}22.12\%\\ (1.99\%)\end{tabular}}} & \begin{tabular}[c]{@{}c@{}}50.44\%\\ (7.99\%)\end{tabular} & \begin{tabular}[c]{@{}c@{}}57.47\%\\ (0.40\%)\end{tabular} & \textbf{\begin{tabular}[c]{@{}c@{}}41.18\%\\ (3.67\%)\end{tabular}} \\ \midrule
\textbf{\begin{tabular}[c]{@{}l@{}}Inst/Cycle\\ (avg/stddev)\end{tabular}} & \begin{tabular}[c]{@{}c@{}}0.536\\ (0.06)\end{tabular}     & \begin{tabular}[c]{@{}c@{}}0.511\\ (0.04)\end{tabular}     & \multicolumn{1}{c|}{\textbf{\begin{tabular}[c]{@{}c@{}}1.221\\ (0.10)\end{tabular}}}     & \begin{tabular}[c]{@{}c@{}}0.892\\ (0.08)\end{tabular}     & \begin{tabular}[c]{@{}c@{}}1.362\\ (0.22)\end{tabular}     & \textbf{\begin{tabular}[c]{@{}c@{}}1.65\\ (0.33)\end{tabular}}      \\ \bottomrule
\end{tabular}
\caption{Hardware counters show that pipelining reduces cache misses, which translates to higher performance.}
\label{table:pipelining}
\end{table}

%% file: related.tex
\section{Related Work}
\label{sec:related}

SAs are influenced by work on building new common runtimes or IRs for data
analytics~\cite{yu2008dryadlinq,palkar2017weld,lee2011implementing,sujeeth2014delite}
and machine learning~\cite{sujeeth2011optiml,abadi2016tensorflow,tvm}.
Weld~\cite{palkar2017weld} and Delite~\cite{sujeeth2014delite} are two specific
examples of systems that use a common IR to detect \emph{parallel patterns} and
automatically generate parallel code. Although \sn does not generate code, we
show in \S\ref{sec:evaluation} that in a parallel setting, the most
impactful optimizations are the data movement ones, so SAs can achieve
competitive performance without requiring developers to replace code.
API-compatible replacements for existing libraries such as
Bohrium~\cite{kristensen2014bohrium} also have completely re-engineered
backends.

Several existing works provide black-box optimizations and automatic
parallelization of functions. Numba~\cite{numba} JITs code using a single
decorator, while Pydron~\cite{muller2014pydron}, Dask~\cite{dask} and
Ray~\cite{ray} automatically parallelize Python code for multi-cores and
clusters. In C, frameworks such as Cilk~\cite{cilk} and
OpenMP~\cite{dagum1998openmp} parallelize loops using an annotation-style
interface. Unlike these systems, in addition to parallelization, SAs enable data
movement optimizations across functions and reason about pipelining safety.

The optimizations that SAs enable have been studied before:
Vectorwise~\cite{zukowski2012vectorwise} and other vectorized
databases~\cite{lang2016data,crotty2015tupleware, boncz2005monetdb} apply the
same pipelining and parallelization techniques as SAs for improved cache
locality. Unlike these databases, \sn applies these techniques on a diverse set
of black-box libraries and also reason about the \emph{safety} of pipelining
different functions using split types. SAs are also influenced by prior work in
the programming languages community on automatic loop
tiling~\cite{grosser2011polly},
pipelining~\cite{grelck2005loop,chakravarty2007data}, and link-time
optimization~\cite{fernandez1995simple,bala2011dynamo}, though we found these
optimizations most effective over nested C loops in user code, and not over compositions of
complex arbitrary functions.

Finally, split types are conceptually related to Spark's
partitioners~\cite{spark-partitioner} and Scala's parallel collections
API~\cite{prokopec2011generic}. Scala's parallel collections API in particular
features a \code{Splitter} and \code{Combiner} that partition and aggregate a
data type, respectively. Unlike this API, SAs enable pipelining and also reason
about the safety of pipelining black-box functions: Scala's collections API still
requires introspecting collection implementations. Spark's
Partitioners similarly do not enable pipelining. 

%% file: conclusion.tex
\section{Conclusion}
\label{sec:conclusion}
Data movement is a significant bottleneck for data-intensive applications that
compose functions from existing libraries. Although researchers have developed
compilers and runtimes that apply data movement optimizations on existing
workflows, they often require intrusive changes to the libraries themselves. We
introduced a new black-box approach called split annotations (SAs), which
specify how to safely split data and \emph{pipeline} it through a parallel
computation to reduce data movement. We showed that SAs require no changes to
existing functions, are easy to integrate, and provide performance
competitive with clean-slate approaches in many cases.

%% file: acknowledgements.tex
\section{Acknowledgements}
We thank Paroma Varma, Deepak Narayanan, Saachi Jain, Aurojit Panda, and the members of the DAWN project for their input on earlier drafts of this work. We also thank our shepherd, Raluca Ada Popa, and the anonymous SOSP reviewers for their invaluable feedback. This research was supported in part by supporters of the Stanford DAWN project (Ant Financial, Facebook, Google, Infosys, Intel, Microsoft, NEC, SAP, Teradata, and VMware), by Amazon Web Services, Cisco, and by NSF CAREER grant CNS-1651570. Any opinions, findings, and conclusions or recommendations expressed in this material are those of the author(s) and do not necessarily reflect the views of the National Science Foundation.